\documentclass[superscriptaddress,twocolumn,showpacs]{revtex4}

\usepackage{graphicx}
\usepackage{dcolumn}

\newcommand{\bb}{\begin{equation}}
\newcommand{\en}{\end{equation}}

\begin{document}

\title{Effects of counterion fluctuations in a polyelectrolyte brush}

\author{C.D. Santangelo}
\affiliation{Department of Physics, University of California,
Santa Barbara, CA 93106, USA}
\author{A.W.C. Lau}
\affiliation{Department of Physics and Astronomy,
University of Pennsylvania, Philadelphia, PA 19104, USA}

\date{\today}

\begin{abstract}
We investigate the effect of counterion fluctuations in a single
polyelectrolyte brush in the absence of added salt by
systematically expanding the counterion free energy about
Poisson-Boltzmann mean field theory.  We find that for strongly charged brushes,
there is a collapse regime in which the brush height decreases with increasing
charge on the polyelectrolyte chains.  The transition to this collapsed regime is
similar to the liquid-gas transition, which has a first-order line terminating
at a critical point.  We find that for monovalent counterions the
transition is discontinuous in theta solvent, while for multivalent counterions
the transition is generally continuous.  For collapsed brushes, the brush height is not independent
of grafting density as it is for osmotic brushes, but scales linear with it.
\end{abstract}
\maketitle

\section{Introduction}
\label{intro}

Recently, correlation effects in electrostatics of highly charged
macroions in aqueous solution have received a
great deal of attention (see Refs.\,\cite{review1,review2,review3}, for recent reviews).
These effects give rise to interesting phenomena that are not contained in
the mean-field Poisson-Boltzmann (PB) theory \cite{contactvalue}, such as
charge inversion \cite{review1} and like-charge attraction \cite{review2,review3}.
One plausible origin of the attraction comes from fluctuations of the
counterion density that become correlated leading to a long-range
attraction that is similar to van der Waals forces~\cite{fluct1,fluct2,andy}.
Counterion fluctuations might have important effects on the conformation of
highly-charged polyelectrolytes \cite{ramin,schiessel}.
In this paper, we study the effects of counterion
fluctuations in a polyelectrolyte brush, where these effects
are thus expected to be important.

\begin{figure}[bp]
{\par\centering
\resizebox*{3in}{2.1in}{\rotatebox{0}{\includegraphics{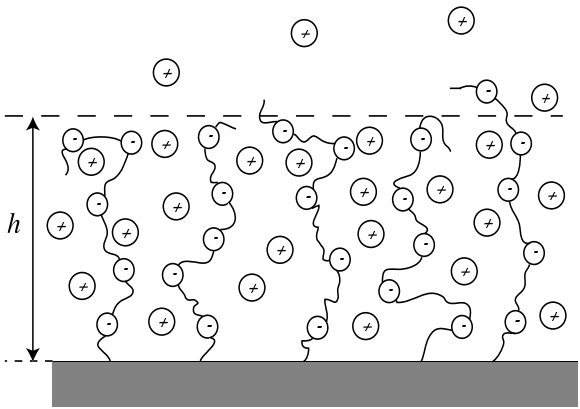}}}
\par}
\caption{A schematic picture of a polyelectrolyte brush. }
\label{brush}
\end{figure}

A polyelectrolyte brush consists of a high density of charged polymers with
one of their ends grafted to a surface (see Fig.\ref{brush}).  Because of their
technological applications, such as enhanced stabilization against colloidal
flocculation, they have been extensively studied in the past both theoretically~
\cite{brushtheory1,brushtheory2,fylbrush,fylpoor,brushtheory3,brushtheory3.5,brushtheory4}
and experimentally~\cite{jlb,brush-expt1,brush-expt2,brush-expt3}.  Pincus \cite{fylbrush}
studied polyelectrolyte brushes, based on the
Alexander-deGennes approximation \cite{alexdegennes}, where monomer density is assumed to be
constant up to the brush height, and mean-field PB theory for the electrostatic
interaction between the charged monomers and the counterions.  The brush height, a quantity that
is accessible to experiments, is then determined self-consistently by minimizing
the total free energy of the brush.  In a theta solvent,
it simply consists of the elastic energy arising from stretching the chains and the
electrostatic (mean-field) free energy.  In this picture, there are basically two regimes --
weak-charging and strong-charging (osmotic) regimes.  In the weak-charging regimes, the
counterions are diffuse and the brush height scales with the degree of polymerization of the
charged polymer $N$, like $h \sim N^3$.  In the osmotic
regime, most of the counterions are trapped inside the brush, whose osmotic pressure
(which tends to increase the brush height) is balanced by elasticity of the chains
(which tends to decrease the brush height).  In this limit, the chains are stretched and
the brush height, $h \sim N$ and is independent of the grafting density.  In later
studies, this mean-field electrostatic picture has been extended to
poor solvent, where a first-order collapse transition was predicted \cite{fylpoor},
to the quasi-neutral regime where excluded volume effects are important \cite{brushtheory3},
and to the self-consistent field theory which goes beyond the Alexander-deGennes
approximation \cite{brushtheory3.5}.

Early experiments on moderately charged polyelectrolyte brush show that
the above theory is reasonably valid \cite{jlb,brush-expt1,brush-expt2,brush-expt3}.
However, recent experiments on highly charged brushes have shown behavior that cannot be
described within mean-field PB theory, especially
in the presence of multivalent counterions.  Experiments on the normal
forces between two brushes~\cite{tirrell} have seen decreases of brush height with
multivalent salt and condensation beyond that predicted by Manning theory.  There
is also evidence of attractive interactions between the brushes~\cite{Schorr}.  In
addition, Bendejacq \textit{et al.}~\cite{denis} have observed a regime in highly
charged cylindrical brushes of annealed polyelectrolytes characterized by a continuous
decrease in brush height with increasing brush charge.  The nonmonotonic behavior of
the brush height occurs even with monovalent counterions and there exists a novel
scaling relation between the brush height and the concentration of added salt.
Finally, a recent simulation~\cite{seidel} also finds enhanced condensation and
a collapsed brush regime even with monovalent counterions. It also shows for large
values of grafting density, the brush height does not follow any of the scaling laws
above.

These observations strongly suggest that counterion correlation
effects are important in highly charged brushes, especially in the
presence of multivalent ions.  In Ref.~\cite{scalingtheory}, Csajka {\em et al}.\
developed a scaling theory describing strongly charged brushes using the so-called box model.
In this model, the chains in the brush are assumed to be equally stretched and the counterions are assumed
to be distributed with constant density.  In addition to the mean-field electrostatic
free energy, a counterion fluctuation energy, estimated using the Debye-H\"{u}ckel
free energy of electrolytes \cite{landau}, is added to reflect correlation effects.
This new term tends to decrease the brush height.  This model predicts a collapsed regime
in which the Debye-H\"{u}ckel fluctuation pressure is balanced by the second virial pressure.
The transition between the osmotic brush and the collapsed
brush is predicted to be first order.

In this paper, we develop a theoretical framework that goes beyond scaling theory and
systematically study the role of counterion density fluctuations in
the absence of added salt.  This is done by employing a field theory for
the counterion contribution to the free energy and \emph{systematically}
expanding about the mean field counterion density to second
order in the fluctuations~\cite{netz,andy}.  The chains are treated within
the Alexander-deGennes approximation of equal stretching and constant monomer density.
We find that for strongly charged brushes, there is a collapse
transition in which the brush height decreases with increasing
charge on the polyelectrolyte chains.  Unlike the scaling theory \cite{scalingtheory},
we find that the transition is similar to the liquid-gas transition with a line
of first-order phase transition terminating at a critical point. Surprisingly,
the valence of the counterions plays an important role: the transition
to the collapse regime is discontinuous for monovalent counterions, while it is
continuous for multivalent counterions.  Furthermore, for collapsed brushes,
we find that the brush height is no longer independent of the grafting density, as in
osmotic brushes, but scales linearly with it.  Our results are expected to be relevant to ongoing experimental
studies in polyelectrolyte brushes and our framework can be easily generalized to study
the important case of the interaction between two highly-charged brushes..

The organization of the paper is as follows: In section~\ref{sec:freeenergy},
we review the expansion about the PB mean field theory in order
to establish notation.  In section~\ref{sec:single}, we specialize to the case
of a single polyelectrolyte brush and calculate the phase diagram in both theta
solvent and good solvent.  Finally, in section~\ref{sec:discussion}, we discuss
the meaning and limitations of our results.  The detailed calculation of the
fluctuation effects have been left to the Appendix.

\section{Counterion Free Energy}
\label{sec:freeenergy}

In this section, we briefly review the theoretical method that allows us
to compute in a rigorous fashion the total electrostatic free energy that
includes fluctuation contributions from the counterions.  The details can be
found elsewhere~\cite{condensation,netz}.  The idea is to map the partition
function of a system of counterions interacting among themselves and with a fixed
distribution of charges to a field theory characterized by an effective Hamiltonian.
We can then systematically expand about its saddle-point solution,
which corresponds to nothing but the mean-field PB approximation,
and calculate the fluctuation contributions to the lowest order.
While the discussion here is general, we will apply this formulation to our model
of polyelectrolyte brush in Sec. \ref{sec:single}.

Consider the primitive model where a system of $N_c$ point-like charged
particles of charge $ - Z e$ (where $Z$ is the valence) and external fixed charges with
charge density $\sigma({\bf x})= e n_f({\bf x})$, immersed in an aqueous
solution with dielectric constant $\epsilon$. The
electrostatic energy for this charged system is \bb
\beta E_{N_c} = Z^2 l_B \sum_{j
> k}^{N_c} { 1 \over \mid {\bf x}_j - {\bf x}_k \mid} -
\sum_{j=1}^{N_c}\, \phi({\bf x}_j),
\label{energy}
\en
where
$\phi({\bf x}) = Z l_B \int d^3{\bf x}'\,{n_f({\bf x}') \over \mid
{\bf x} - {\bf x}' \mid}$ is the ``external'' electric potential from the
fixed charge distribution and $l_B \equiv {e^{2} \over \epsilon k_{B}T} \approx 7\,$\AA$\,\,$
is the Bjerrum length in water at room temperature,
$k_B$ is the Boltzmann constant, and $T$ is the temperature.   The partition
function is\bb
Z_{N_c}[\phi] = {1 \over
N_c!}\,\prod_{i=1}^{N_c}\,\int\,{d^3{\bf x}_i \over a^3}\,
\exp{(-\beta E_{N_c})}, \label{part}
\en
where $a$ is the molecular size
of the counterions and $\beta \equiv 1/(k_B T)$.  The partition function of Eq.\ (\ref{part}) can be
mapped into a field theory, by switching to the grand canonical ensemble.
We introduce a chemical potential $\mu$ (in units of $k_B T$) and express
the grand partition function as\bb
{\cal Z}_{\mu}[\phi] = \sum_{N_c =0}^{\infty}\,
e^{\mu N_c}\,Z_{N_c}[\phi].
\en
After a Hubbard-Stratonovich transformation, we obtain the
functional representation for the system of charges in an external field:
\begin{eqnarray}
{\cal Z}_{\mu}[\phi] &=& {\cal N}_0\,\int {\cal D}\psi\,
e^{- {\cal S}[\psi,\phi]},
\label{partition} \\
{\cal S}[\psi,\phi] &=& \int { d^3{\bf x} \over \ell_B }\,\left \{
{1 \over 2}\psi({\bf x})[-\nabla^2]\psi({\bf x}) -
\kappa^2e^{i \psi({\bf x}) + \phi({\bf x})}\right \} \nonumber \\
&\equiv& \int d^3{\bf x}\,{\cal H}[\psi,\phi],
\label{action}
\end{eqnarray}
where $\ell_B = 4 \pi l_B Z^2$,
$\kappa^2 = \rho_0 \ell_B$, $\rho_0 = {e^{\mu} /
a^3}$, and ${\cal N}_0^{-2} \equiv \det[-\nabla_{\bf x}^2]$ is
the normalization factor.  The saddle point, $\psi_0({\bf x})$,
given by $\left. { \delta {\cal S} \over \delta \psi({\bf x}) }\right |_{\psi =\psi_0} = 0$,
is determined by the equation \bb
\nabla^2[i\psi_{0}({\bf x})] = \kappa^2
e^{i \psi_{0}({\bf x}) + \phi({\bf x})},
\label{saddle}
\en
which after introducing the mean-field potential,
$ \varphi({\bf x})=  - i \psi_{0}({\bf x})
- \phi({\bf x})$, becomes \begin{eqnarray}
\nabla^2\varphi({\bf x}) + \kappa^2e^{-\varphi({\bf x})} &=& -
\nabla^2 \phi({\bf x}) \nonumber \\
&=& { \ell_B \over Z}\,n_f({\bf x}),
\label{pbeqn}
\end{eqnarray}
which is the Poisson-Boltzmann equation.  This equation will be
examined more closely in Sec.\ \ref{subsec:meanfieldtheory} for our
model of polyelectrolyte brush to be introduced in Sec.\ \ref{sec:single}.

To obtain the free energy for the counterions at the mean-field level,
we note that it is related to the Gibbs potential
$\Gamma_0[\phi] \equiv  {\cal S}[\psi_0, \phi]$ by a
Legendre transformation:\bb
\beta F_0(n) =  \Gamma_0[\phi] + \mu \int d^3{\bf x}\,
\rho_0({\bf x}),
\label{mean}
\en
where $\rho_0({\bf x})$ is the mean-field free counterion density given
by\bb
\rho_0({\bf x}) = \rho_0 e^{i \psi_0({\bf x}) + \phi({\bf x})\,}.
\en
To capture correlation effects, we expand the action ${\cal S}[\psi,\phi]$
about the saddle-point, $\psi_0({\bf x})$, to second order in $\Delta \psi({\bf x}) =
\psi({\bf x}) -\psi_0({\bf x})$:\begin{eqnarray}
{\cal S}[\phi,\psi] &=& {\cal S}[\phi,\psi_0] \\
&+& {1 \over 2}\,\int d^3{\bf x}\,\int d^3{\bf y}\,
\Delta\psi({\bf x})\,\hat{{\bf K}}({\bf x}, {\bf y})\,\Delta\psi({\bf y})+ \cdots,
\nonumber
\end{eqnarray}
where the differential operator\bb
\hat{{\bf K}}({\bf x}, {\bf y}) \equiv
\left [ - \nabla_{{\bf x}}^2\,+ \,
\kappa^2 e^{i \psi_0({\bf x}) + \phi({\bf x})\,} \right ] \delta( {\bf x} - {\bf
y}),
\label{K}
\en
is the second variation of the action ${\cal S}[\psi,\phi]$.  Notice
that the linear term in $\Delta \psi({\bf x})$ does not contribute to the expansion
since $\psi_0({\bf x})$ satisfies the saddle-point equation Eq.\ (\ref{saddle}).
Performing the Gaussian integrals in Eq.\ (\ref{partition}), we obtain an
expression for the change in the free energy due to fluctuations of the
counterions:\bb
\beta \Delta {\cal F} = {1 \over 2} \ln \det \hat{{\bf K}}  -
{1 \over 2} \ln \det [ -\nabla_{\bf x}^2] ,
\label{3D}
\en
where the second term comes from the normalization factor ${\cal N}_0$.
Thus, combining Eqs. (\ref{mean}) and (\ref{3D}) together,
the total electrostatic free energy of the system can be expressed as\bb
\beta F_{el} = \beta F_0 + {1 \over 2} \ln \det \hat{{\bf K}}
- {1 \over 2} \ln \det[ -\nabla_{\bf x}^2].
\label{totalfree}
\en
Thus, the total electrostatic free energy is comprised of a mean-field
free energy and a fluctuating free energy.  Note that the expansion parameter
of the total free energy is proportional $\kappa\,\ell_B$, where $1/\kappa$
roughly corresponds to the average distance between charges. Thus, for highly charged systems,
the fluctuating term, linear in $\kappa\,\ell_B$, may become important,
and the mean-field PB free energy no longer provides a reasonable
approximation.  The computation of the fluctuating free energy
relies on diagonalizing the operator $\hat{{\bf K}}({\bf x}, {\bf y})$,
which depends on the mean-field solution.   For counterions in a uniform background,
this term gives the usual Debye-H\"{u}ckel free energy,
$\Delta F / V= - k_B T \kappa^3/(12 \pi)$ \cite{landau}. However, it might be
difficult to calculate this term in general, and exact evaluation is known
only for planar systems \cite{condensation,andy}.  Indeed, as we will
see below, an approximation must be introduced in order to compute this
term analytically for our model of polyelectrolyte brush.

\section{Single Polyelectrolyte Brush}
\label{sec:single}

We consider a polyelectrolyte brush (see Fig.\ \ref{brush}),
formed by a long intrinsically flexible charged polymers grafted
at one end onto an impermeable planar surface and immersed in a salt-free
solution.  We consider the simple case of monodisperse polymers of
degree of polymerization, $N$, of which a fraction $f$ is charged, with
grafting density $\rho$.  We avoid the complicated issue of how electrostatics
induces stiffness of the chains, and simply assume that the polymers
are Gaussian chains, with Kuhn length $a$ which is the same
molecular size of the counterions.

In addition to the electrostatic free energy, we have to model the energetics of
the polyelectrolyte brush, arising from elasticity and its interaction with the solvent.
Within the Alexander-deGennes approximation \cite{alexdegennes}, in which
monomer density are assumed to be constant in a region of height $h$,
all the chains are stretched by an equal amount and store
an elastic energy per unit area of\bb
\beta F_G /{\cal A}=  { 3\,h^2 \over 2 N a^2 }\rho,
\label{elastic}
\en
where ${\cal A}$ is the area of the surface.  The solvent quality is
described in terms of the Flory-Huggins free energy per unit area \cite{deGennes}
\bb
\beta F_V/{\cal A} = \frac{v}{2} a^3 c^2 h + \frac{w}{6} a^6 c^3 h.
\en
where $c = N \rho/h$ is the average monomer density, $v$ is the
dimensionless excluded-volume parameter, which is positive for good solvents
and negative for poor solvent, and $w$ is the third
virial coefficient which is typically positive and of order unity.
In the next subsection, we discuss the electrostatic free energy within
our model of polyelectrolyte model, as outlined in Sec.\ \ref{sec:freeenergy}.
In Sec.\ \ref{subsec:phasediagram}, we present the phase diagram of our
model.

\subsection{Mean-field theory and fluctuation contribution}
\label{subsec:meanfieldtheory}

The fixed charge distribution for an Alexander-de Gennes
polyelectrolyte brush is $n_f({\bf x}) = n_0 \tilde{\Theta}(z-h)$,
where $n_0 = f c= N f \rho /h$ and $f$ is the fraction of monomers that are
charged. The function $\tilde{\Theta}(x) = 1- \Theta(x)$, where
$\Theta(x)$ is the Heaviside step function, which equals $1$ when
$x>0$ and $0$ when $x<0$. Due to translation invariance in the
$xy$-plane, the PB equation, Eq.\,(\ref{pbeqn}) can be written as
\begin{eqnarray}
{d^2 \varphi(z) \over dz^2 } + \kappa^2e^{-\varphi(z)}
= { \ell_B n_0\over Z}\tilde{\Theta}(z-h).
\label{pbbrush}
\end{eqnarray}
Outside the brush ($z>h$), the right-hand side of Eq.\ (\ref{pbbrush}) is
zero and its solution is the same as that of a charged plane
held at a constant potential:\bb
\varphi_>(z) = \varphi_h + 2 \ln\left [ 1 + {z -h \over \lambda} \right ],\,\,z > h,
\en
where $\varphi_h$, which will be determined below, is the potential
at the brush height and $\lambda$ is the Gouy-Chapmann length, determined by
$\lambda = {\sqrt{2}\,e^{\varphi_h/2} / \kappa}$, which
is a measure of the extent of how diffuse the counterion distribution is
outside of the brush.

\begin{figure}[bp]
{\par\centering
\resizebox*{3in}{2.1in}{\rotatebox{0}{\includegraphics{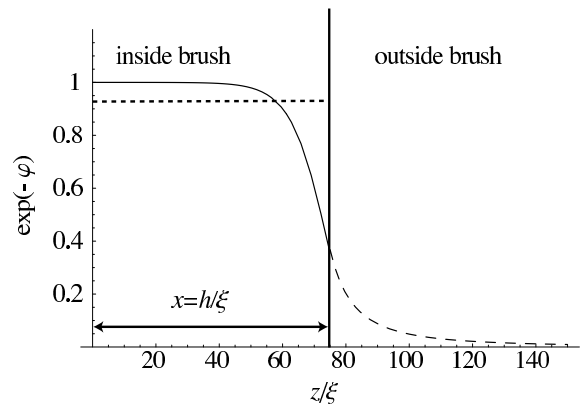}}}
\par}
\caption{A plot of the mean-field counterion density in the osmotic brush limit.
It is clear that most of the counterion is trapped inside the brush.
The dashed line represents the spatially averaged counterion density inside the
brush [See, Eq.\ (\ref{averaged})].  This approximation is used to compute
the fluctuation free energy.} \label{MFcounteriondensity}
\end{figure}

Inside the brush we must solve\bb
{d^2 \varphi_< \over dz^2 } + \kappa^2e^{-\varphi_<}
= { \ell_B n_0\over Z}, \,\, z<h.
\en
Multiplying this equation with $\varphi_<$, integrating once,
and imposing the boundary conditions that the potential and the
electric field at the grafting surface be zero, ``the
constant of motion" is found to be $-\kappa^2$, and we have\bb
\label{eq:firstintegral}
{1 \over 2} \left ( {d \varphi_< \over dz } \right )^2  =
\kappa^2 \left [ e^{-\varphi_<} - 1 \right ] + { \ell_B n_0 \varphi_< \over Z}.
\en
Using the continuity of the electric field and potential across
the boundary at $z=h$, we find that $\kappa^2$ is related to $\varphi_h$ by \bb
\kappa^2 = { \ell_B n_0 \varphi_h \over Z} =\frac {2 \varphi_h}{\xi h},
\en
where we have defined $\xi \equiv 2 Z / (\ell_B n_0 h)$. Note that this length is
independent of $h$. The electrostatic potential inside the brush, $\phi_<(z)$,
is determined implicitly by\bb
\int_0^{\varphi_<}{d\varphi ' \over \sqrt{ e^{-\varphi '} -1 + {\varphi '\over \varphi_h}}}
\equiv \Delta[ \varphi_<(z), \varphi_h] = 2 \sqrt{\frac{\varphi_h}{\xi h}} z.
\label{insidebrush}
\en
In particular, the potential at the brush height, $\varphi_h$ is determined
by the relation\bb
\label{eq:reln}
\Delta[ \varphi_h, \varphi_h] = 2 \sqrt{\varphi_h \frac{h}{\xi}}.
\en
In the weak charging limit, it gives $\varphi_h \approx h/\xi \ll 1$, and in the strong
charging limit, $\varphi_h \approx 1$.  Fig.\ \ref{MFcounteriondensity} plots
the counterion density.  It is
clear in the strong charging limit that while most of the counterions is in the brush,
there are still some outside.  This point will be crucial when we discuss the fluctuation
contributions below.

\begin{figure}[bp]
{\par\centering
\resizebox*{3in}{2.1in}{\rotatebox{0}{\includegraphics{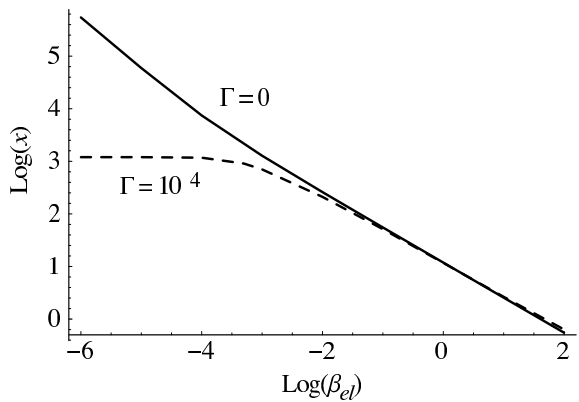}}}
\par}
\caption{The height versus beta for values of $\Gamma =0, 10^4$.
} \label{xvsbeta}
\end{figure}

The mean-field electrostatic free energy per unit area can be
calculated straightforwardly using the solution described above and Eq.\ (\ref{mean}).
The result is\begin{eqnarray}
\beta f_0 & = &  \frac{N f \rho}{Z} \ln
\left( \frac{2 \varphi_h a^3}{\ell_B \xi h}  \right) -
\frac{2 \varphi_h}{\ell_B \xi} \\
&-& { 4 \over \ell_B}\sqrt{ {\varphi_h \over \xi h} }\left [\, e^{-{\varphi_h \over 2}}
+ {1\over 2}  \int_0^{\varphi_h}
d\varphi \sqrt{e^{-\varphi} - 1 +{\varphi \over \varphi_h}} \,\right ],
\nonumber
\end{eqnarray}
where we have made used of the fact that the chemical potential
is given by $\mu = \ln \left[ {\varphi_h a^3 /(2 \xi^2 \ell_B)} \right]$.
In the mean-field picture, electrostatics contributes an outward pressure, $
\Pi_{\mbox{\tiny MF}}(h) = - { \partial f_0 / \partial h},$
which scales as $\Pi_{\mbox{\tiny MF}}(h) \sim 1 /( \ell_B \xi h)$, in the strong charging
limit, and $ \Pi_{\mbox{\tiny MF}}(h) \sim 1 /(  \ell_B \xi^2)$ in the weak charging limit.
Note that this mean-field pressure is always repulsive.  Balance this electrostatic
pressure against that arising from the elasticity
of Eq.\ (\ref{elastic}), we have $h \approx \xi/ \beta_{el}$, for $\beta_{el} \ll 1$
(strong charging) and $h \approx \xi/ \beta_{el}^2$, for $\beta_{el} \gg 1$
(weak charging), where we have defined an important dimensionless parameter
$\beta_{el} \equiv {Z^{1/2} \xi/ (N f^{1/2} a)}$, which characterizes the importance
of the elastic term compared to electrostatic pressure.  These scaling regimes
(as plotted in Fig.\ \ref{xvsbeta}) are well-known as discussed in Ref.\ \cite{fylbrush}.
Here, we study deviations from these scaling laws when counterion fluctuations
are taken into account.

To evaluate the fluctuation contribution to the free energy,
we start with the expression for the change in the free energy due to
fluctuations of the counterions, Eq. (\ref{3D})
\begin{eqnarray}
\beta \Delta {\cal F} =  { 1\over 2}\ln \det \hat{{\bf
K}} - {1 \over 2} \ln \det \left [ - { \nabla^2 } \right ],
\nonumber
\end{eqnarray}
where the operator $\hat{{\bf K}}$ for our model of
polyelectrolyte brush is\begin{eqnarray}
{\bf K}({\bf x},{\bf y}) &\equiv& \,\left [\, - \nabla^2_{\bf x}
+ \kappa^2 e^{-\varphi_<}\, \tilde{\Theta}(z -h)
\vphantom{+ {2 \Theta(z -h) \over ( z- h +\lambda)^2 }}
\right. \nonumber\\
&&  \hphantom{- \nabla^2_{\bf x} + \kappa }
+ \left. {2 \Theta(z -h) \over ( z- h +\lambda)^2 } \right ] \delta( {\bf x} -{\bf y}).
\end{eqnarray}
The derivative of $\Delta {\cal F}$ with respect to $h$ can be straightforwardly
calculated by making use of the exact identity
$\delta \ln \det \hat{{\bf X}} = {\mbox{Tr}}\,\hat{\bf X}^{-1}\,
\delta\,\hat{\bf X}$: \begin{eqnarray}
\label{partialf}
{ \partial \beta \Delta {\cal F} \over \partial h }
&=& { 1 \over 2 \ell_B} \int d^{3}{\bf x}\, G({\bf x},{\bf x})\,\\
&\times& {\partial \over \partial h }\, \left [ \kappa^2 e^{-\varphi_<}\, \tilde{\Theta}(z -h)
+ {2 \Theta(z -h) \over ( z- h +\lambda)^2 }  \right ], \nonumber
\end{eqnarray}
where $G({\bf x},{\bf x})$ is the Green's function (inverse) of
${\bf K}({\bf x},{\bf y})$ satisfying\begin{eqnarray}
&&\left [\,- \nabla^2_{{\bf x}} + \kappa^2 e^{-\varphi_<}\, \tilde{\Theta}(z -h)
+ {2 \Theta(z -h) \over ( z- h +\lambda)^2 } \,\right ] G({\bf x},{\bf x}')\nonumber \\
&& \hphantom{2 \Theta(z -h)\,- \nabla^2_{{\bf x}} + \kappa^2 e^{-\varphi_<}
\, \tilde{\Theta}(z)  }
  = \ell_B\,\delta( {\bf x} - {\bf x}').
\end{eqnarray}
Unfortunately, $G({\bf x},{\bf x}')$ cannot be evaluated analytically because the
counterion density inside the brush is not an explicit function of $z$
[see Eq.\ (\ref{insidebrush})].  In order to make analytical progress, we employ the reasonable
approximation that the counterion density inside the brush is a constant averaged
over the brush, given by \bb
\Lambda^2 \equiv \frac{1}{h} \int_0^h dz \kappa^2 e^{-\varphi_<(z)}
= { 2 \over \xi h  }\left ( 1  - { \xi \over \lambda } \right ).
\label{averaged}
\en
This approximation entails neglecting the spatial variations inside the
brush. Judging from Fig.\ \ref{MFcounteriondensity}, this
approximation should not be in serious error.

\begin{figure}
    \label{fig:pressurefig2}
    \resizebox{3in}{!}{\includegraphics{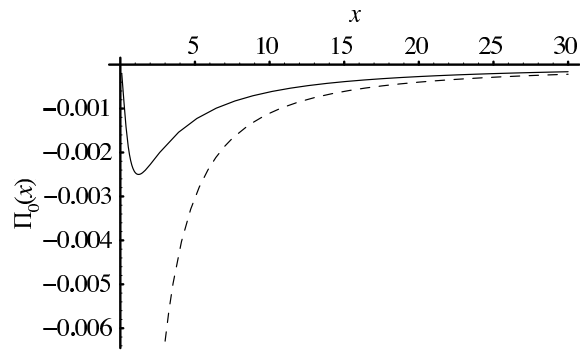}}
    \caption{\label{fig:pressurefig}
    The fluctuation contribution to the free energy, $\Pi_0(x)$ (solid line),
    where $x = h/\xi$ is the rescaled height variable.  The scaling prediction
    for the fluctuation pressure from equation (\ref{eq:scalingprediction})
    (dashed line) diverges as small $x$.}
\end{figure}

Within this approximation, $G({\bf x}, {\bf x'})$
can be evaluated analytically (see Appendix), and we find\begin{eqnarray}
\label{eq:pressure}
{ 1 \over {\cal A}}\,{ \partial \beta \Delta {\cal F} \over \partial h }
& = & {\Lambda^2 \over h } \left [\,{\cal I}_1(h) - {\cal I}_2(h)\,\right ]
+ { {\cal I}_5(h) \over h^3} \nonumber\\
& + & { 1 \over \lambda^2 h }{\partial \lambda \over \partial h}
\left [\,{\cal I}_3(h) - {\cal I}_4(h) + 2\,{\cal I}_2(h)\, \right ],
\end{eqnarray}
where the dimensionless integrals ${\cal I}_i(h)$ are defined in the Appendix.
This is the main result of this paper.  It says that counterion
fluctuations contribute an inward (negative) pressure $\Pi_{f}(h)
= -\,{ {\cal A}^{-1}}\,{ \partial \Delta {\cal F} /\partial h }
\equiv \Gamma \,\xi^{-3}\,\Pi_0(h/\xi)$, as plotted in Fig.\ \ref{fig:pressurefig}.
First, the fact that it is negative is consistent with the notion that
the fluctuations must lower the free energy.  The strength of this pressure is controlled by
$\Gamma = \ell_B/\xi$, which roughly scales as $(\ell_B f \rho)^2$. Thus, this term becomes important
for highly charged polyelectrolytes, as expected.

Although the expression Eq.\ (\ref{eq:pressure}) looks rather complicated,
we can understand its meaning by expanding $\Pi_0(h/\xi)$ in the strong charging limit:
\begin{eqnarray}
\label{eq:scalingprediction}
\Pi_0(h/\xi)  &=&  - \,\alpha_1\,\left ({\xi /h}\right )^{3/2} +
\alpha_2\,\left ( {\xi / h} \right )^2 \nonumber \\
&-& \alpha_3 \,\left ( {\xi / h} \right )^{5/2} + \cdots,
\label{expansion}
\end{eqnarray}
where $\alpha_1 \sim 0.0375$ and $\alpha_2 \sim 0.00833$ are constants.
The first term is the fluctuation contribution to the free energy of a
coulomb gas of counterions distributed uniformly throughout the brush volume.
To see this, recall that in the strong charging limit most of the counterions is within the
brush, with a 3D concentration of $c \sim n_0/h$. This implies that the inverse
of the 3D ``screening" length $\kappa_s \sim \sqrt{\ell_B n_0 /h}$.  Using
the Debye-H\"{u}ckel free energy (per unit volume) $\beta \Delta f  \sim - \kappa_s^3$,
we have $\Pi_{f}(h) \sim \partial \left (\Delta f \cdot h \right )/\partial h \sim h^{-3/2}$,
which is the first term in Eq.\ (\ref{expansion}).
In fact, similar argument has been used by Csajka {\em et al}.\ \cite{scalingtheory}
to derive the scaling laws.  The second term in Eq.\ (\ref{eq:scalingprediction})
may be interpret as an effective second virial term arising from the
coupling between fluctuations of the counterions inside
the brush and outside.  Note that this term is not included in the
scaling theory for collapsed brushes.  It is interesting to note that the
divergence at small $h$ of the asymptotic fluctuation pressure is
purely an artifact of the approximation, and it is not present in Eq.\ (\ref{eq:pressure}).
It is likely then that scaling laws derived for the collapsed state of
a polyelectrolyte brush using similar approximate expressions may
not capture the full behavior of the brush.

\subsection{Phase Diagram}
\label{subsec:phasediagram}

Now, we are in the position to discuss the equilibrium phase diagram of the
polyelectrolyte brush.  The important quantity, the equilibrium height of the
brush, is determined by minimizing the total free energy
of the polyelectrolyte brush, $F = F_G + F_V + F_{el}$,
with respect to the brush height.   It is more convenient to employ
the following dimensionless variables:
\begin{eqnarray}
\omega & \equiv & N \rho a^2\\
\nu & \equiv  & \frac{v a^3 Z^2}{\ell_B \xi^2 f^2}.
\end{eqnarray}
in addition to $x \equiv h/\xi $, $\Gamma \equiv \ell_B /\xi$, and
$\beta_{el} \equiv {Z^{1/2} \xi/ (N f^{1/2} a)}$ already defined in the previous section.
The variable $\nu$ is a measure of the contribution to the pressure from the second virial
coefficient compared to the mean field pressure.  Note that the mean-field pressure
and the fluctuation pressure depends on $x$ only.

\begin{figure}
    \label{fig:phasediagramtheta2}
    \resizebox{3in}{!}{\includegraphics{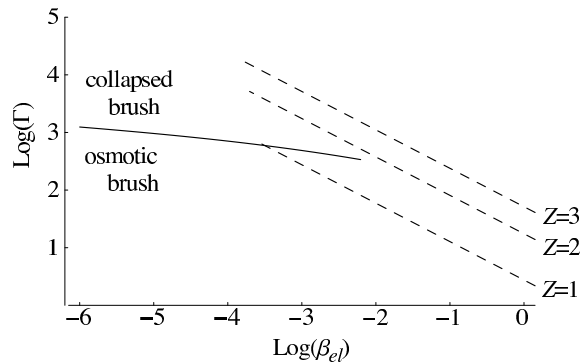}}
    \caption{\label{fig:phasediagramtheta}Typical phase diagram of a
    single brush in theta solvent (in this case, $N=200$, $\omega=1$).
    The solid line represents a first order phase transition between the
    osmotic brush and the collapsed brush regimes.  The Pincus brush regime
    occurs for $\beta \approx 1$ and is not shown.  The dashed curves are
    generated holding all parameters constant except the brush charge
    fraction (arrows indicate increasing charge fraction) for monovalent,
    divalent, and trivalent counterions.}
\end{figure}

\begin{figure}
    \label{fig:heightZ2theta2}
    \resizebox{3in}{!}{\includegraphics{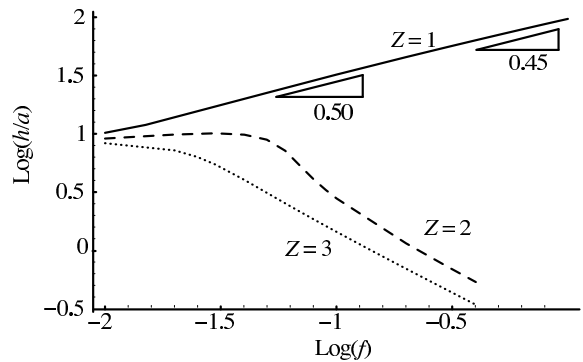}}
    \caption{\label{fig:heightZ2theta} Brush height for monovalent (solid),
    divalent (dashed), and trivalent (dotted) counterions in a theta solvent
    (see figure~\ref{fig:phasediagramtheta}).  The monovalent salt height
    obeys the power law $h \propto f^{0.50}$, though it shows minor
    deviations from power at very strong charging (with an exponent $\approx 0.45$).}
\end{figure}

In Fig.~\ref{fig:phasediagramtheta}, we display the typical phase diagram of a
polyelectrolyte brush in contact with a theta solvent as a function of $\beta_{el}$ and
$\Gamma$.  We find a first order collapse transition line, from the osmotic brush
to the collapse brush, represented by the solid line in Fig.~\ref{fig:phasediagramtheta}.
This line terminates at a critical point, which is well within the strong charging regime,
$\beta_{el} <  1$.  Thus, it is similar to the liquid-gas transition.  This behavior resembles
to that of a polyelectrolyte brush in a poor solvent, where attractive interaction between
monomers mediated by the solvent acts against the electrostatic repulsion to collapse
the brush \cite{fylpoor}. Here, fluctuations induce an attractive interaction
which collapse the brush.  This effect on the brush height can be
clearly seen in Fig.\ \ref{xvsbeta}. It is interesting to note that the valence of the counterions
plays a crucial role.  For typical brushes, increasing the charge fraction while holding other
parameters fixed (as shown with dotted lines in Fig.\ \ref{fig:phasediagramtheta}),
only the monovalent counterion goes through the first-order collapse.
For divalent and trivalent counterions, on the other hand, we observe a
smooth continuous decrease in $h$, as shown in Fig.~\ref{fig:heightZ2theta},
where we plot the brush height as a function of the charge fraction $f$.  As
expected, correlation effects are more important for higher valence $Z$ of the
counterions.  Even for a moderately charge fraction $f$, the brush height already starts to decrease
for for divalent and trivalent counterions.  It is also interesting to observe that
for monovalent counterion (before the collapse transition), the brush height obeys
the osmotic brush scaling law of $h \sim f^{1/2}$ only for moderate $f$, which
crosses over to $h \sim f^{0.45}$.  Deep inside the collapsed regime,
the brush height is determined by the fluctuation pressure against the
third virial term, which gives $h \sim \omega^2\,\xi \, /\,\Gamma^{2/3} \sim f^{-5/3}$.
This is not shown in Fig.\ \ref{fig:heightZ2theta}.

\begin{figure}
    \resizebox{3in}{!}{\includegraphics{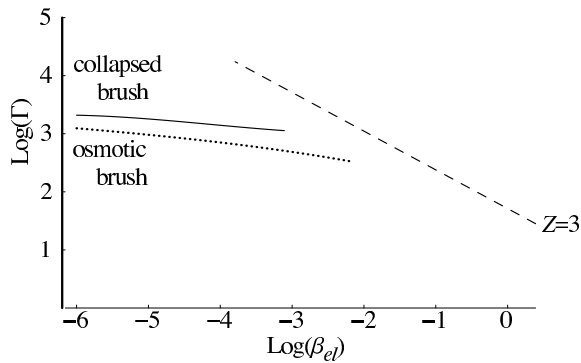}}
    \caption{Typical phase diagram for good solvent ($v = 17.0275$, $Z=3$)
    for the brush of figure~\ref{fig:phasediagramtheta}.  The solid line is a
    first order phase transition from the osmotic to collapsed brush regimes.
    For comparison, the transition for a brush in theta solvent is shown by the
    dotted line.  Notice that in good solvent, the transition line occurs at
    higher values of $\epsilon$ and the critical point occurs for smaller
    values of $\beta_{el}$.  This is also typical of brushes with large
    values of $\omega$.  As a further reference, the dashed line shows the
    curve of increasing charge fraction in for trivalent salt.  For the case
    of monovalent counterions, the first-order transition occurs for smaller
    values of $\epsilon$ than with trivalent counterions.}
    \label{thisisit}
\end{figure}

\begin{figure}
    \label{fig:heightZ12}
    \resizebox{3in}{!}{\includegraphics{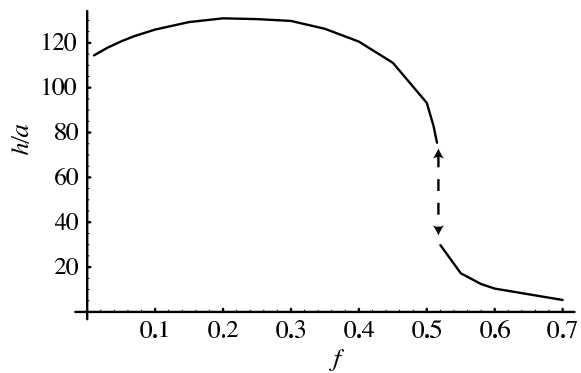}}
    \caption{\label{fig:heightZ1} Brush height vs. charge fraction in good
    solvent for monovalent counterions at high density ($\rho a^2 = 0.0625$).}
\end{figure}

In a good solvent, the polyelectrolyte brushes do not show any major qualitative difference from
a theta solvent.  A typical phase diagram is shown in Fig.\ \ref{thisisit}.
In this case, the first-order phase boundary occurs at higher values of
$\epsilon$ and the critical point shifts to even smaller values of $\beta_{el}$.
This is not surprising because in a good solvent, there is an additional outward
pressure which the fluctuation pressure must overcome before the brush is collapsed.
In addition, it is interesting to observe in Fig.\ \ref{fig:heightZ1}, where we have plotted
the behavior of the brush height as a function of the charge fraction, that
the brush height increases at low charge fraction, then decreases with increasing charge fraction,
and eventually collapse.  This nonmonotonic behavior is qualitatively similar
to that observed by Bendejacq \textit{et al.}~\cite{denis}
for monovalent salt in highly charged and dense cylindrical brushes.

\begin{figure}
    \label{fig:heightvsdensity2}
    \resizebox{3in}{!}{\includegraphics{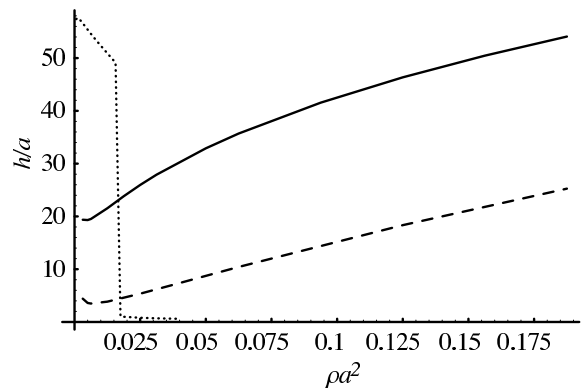}}
    \caption{\label{fig:heightvsdensity} Brush height vs. density ($N=200$, $f=0.3$).
    In theta solvent (dotted curve), the brush height exhibits a first order collapse.
    In good solvent for monovalent (solid) and divalent (dashed) counterions, the brush
    height decreases initially, then increases.  At large densities, the increase is linear.}
\end{figure}

For an osmotic brush at the level of a scaling theory, the brush height is
independent of the grafting density.  In theta solvent, significant deviations
from osmotic brush scaling can be seen even for monovalent counterions as well
as a discontinuous collapse (see dotted curve in Fig.\ \ref{fig:heightvsdensity}).
In good solvent, however, the brush height scales nearly linear in the grafting density,
as shown in Fig.\ \ref{fig:heightvsdensity}.  This scaling can be understood from
balancing of the second virial pressure against the fluctuation pressure:
$h \sim \xi\, \omega^4\,\left [ v\,\xi / (\Gamma a ) \right ]^2 \sim \rho$.

\section{Concluding Remarks}
\label{sec:discussion}

In this paper, we have shown that expanding the counterion free energy systematically about
Poisson-Boltzmann mean field theory for a single brush results in a collapse
transition of the brush height with multivalent counterions.  Though we find a
first order transition similar to that of Csajka {\textit et al.}~\cite{scalingtheory},
we also find a critical point in the phase diagram and find that, generically,
the decrease in brush height is continuous.

The major difference in formulation between our theory and this scaling theory
is that the fluctuations of counterions outside of the brush have been taken
into account at the quadratic level.  Fluctuations of counterions from the
interior to exterior region of the brush are likely to be responsible for
the existence of the critical point in the phase diagram.

Throughout this paper, we have assumed that all chains in the brush are
equally stretched, and that the monomer density is constant throughout the brush.
We have also employed the constant density approximation for the counterion density
within the brush when expanding in the fluctuations.  These approximations have been
necessary in order to calculate the fluctuation pressure, though in reality the
monomer density is known to be approximately parabolic~\cite{borisov}.  It is
possible that the discontinuity of monomer density in our approximation is
responsible for the first order transition region, and that a more exact
theory may have a critical point at much smaller values of $\beta_{el}$ or
no first order transition at all \cite{notran}.

Neglected in our treatment is fluctuations of the polyelectrolyte chains themselves.
These corrections to the free energy are likely to be unimportant if the
fluctuations of uncondensed counterions do not couple strongly to them.
They may play a role, however, when condensed counterions are properly
taken into account.

We have also neglected the inhomogeneities of the monomer density in
the brush, which induces inhomogeneities in the mean field counterion density.
These inhomogeneities are due, in part, to counterion condensation on the
polyelectrolyte chains.  These condensed counterions are not free to fluctuate
and do not contribute to the osmotic pressure due to the counterion entropy.
Further, the condensed counterions can cause additional attractions between
the chains of the brush due to structual correlations beyond those
calculated in this paper.

In future work, we will study the problem of attractive interactions
between two opposing brushes, and work to understand more precisely the
role of condensed counterions in the brush.

\begin{acknowledgements}
We acknowledge discussions with D. Bendejacq, V. Ponsinet,
and in particular T.C. Lubensky and P. Pincus for a critical reading of the
manuscript and indispensable input for this work to take shape.
CDS is supported through the NSF Grant DMR 02-03755 and the MRL Program of the National Science Foundation under Award No. DMR 00-80034.  AWC acknowledges supports
from NSF through the MRSEC Grant DMR 00-79909.
\end{acknowledgements}

\appendix

\section{Calculation of fluctuation energy for a single brush}

Within the constant counterion density approximation within the brush,
the Fourier transform of $G({\bf x}, {\bf x'})$ in the $xy$-plane, $G(z,z';q)$,
satisfies \begin{eqnarray}
&& \left [\,- { \partial ^2 \over \partial z^2} + q^2  + \Lambda^2\, \tilde{\Theta}(z -h)
+ {2 \Theta(z -h) \over ( z- h +\lambda)^2 } \,\right ] G(z,z';q)\nonumber \\
&& \hphantom{2 \Theta(z -h)\,- \nabla^2_{{\bf x}} + \kappa^2 e^{-\varphi_<}
\, \tilde{\Theta}(z)  }= \ell_B\,\delta( z-z'),
\end{eqnarray}
where $\Lambda^2 \equiv 2\left ( 1  - { \xi /\lambda } \right )/(\xi h)$,
and can be solved by standard technique \cite{arfken}.  The result is
\begin{widetext}
\begin{eqnarray}
G_<(z,z;q) &=& {\ell_B \over 2 \alpha}
{ \left [ e^{-\alpha z} - {\cal M}_1(q) e^{+\alpha z} \right ]
\left [ e^{+\alpha z} + {\cal M}_2(q) e^{-\alpha z} \right ]
\over 1 + {\cal M}_1(q) {\cal M}_2(q) },
\hphantom{ 1 + { 1 \over q^2 ( z- h +\lambda)^2}+}\mbox{for}\,\,z<h,\\
G_>(z,z;q) &=& {\ell_B \over 2 q}\left [ 1 -  { 1 \over q^2 ( z- h +\lambda)^2 }\right ]
+ {\ell_B {\cal L}(q) e^{-2q(z-h)} \over 2 q}
\left [ 1 + { 1 \over q( z- h +\lambda)} \right ]^2,\,\,\,\,\,\,\mbox{for}\,\,z>h,\\
{\cal L}(q)   &\equiv& { \left [ 1 - q \lambda ( 1 - q \lambda)\right ]
\left \{ 1 + f_{-}(q) + {\cal M}_2(q) e^{-2 \alpha h} \left [ 1 - f_{-}(q) \right ] \right \}\over
\left [ 1 + q \lambda ( 1 + q \lambda)\right ] \left \{ 1 + f_+(q) +
{\cal M}_2(q) e^{-2 \alpha h} \left [ 1 - f_+(q) \right ] \right \}},
\end{eqnarray}
where $\alpha^2 = q^2 + \Lambda^2$,
${\cal M}_1(q) \equiv e^{-2 \alpha h} \left [{ 1 - f_{+}(q) \over 1+f_{+}(q)}\right ]$,
${\cal M}_2(q) \equiv(\alpha -q )/ (\alpha + q )$, and $f_{\pm} \equiv
\alpha \lambda ( 1 {\pm} q \lambda ) /\left [ 1 {\pm} q \lambda (1 \pm q\lambda)\right].$
Now, returning to the expression of the derivative of
$\Delta {\cal F}$ with respect to $h$, Eq.\ (\ref{partialf}), by noting that
the derivative of ${\partial \Theta(z -h) / \partial h }= - \delta( z - h)$ and
${\partial \tilde{\Theta}(z -h) / \partial h }= + \delta( z - h)$, we find that
Eq.\ (\ref{partialf}) consists of three parts:\bb
{1 \over {\cal A}} { \partial \Delta {\cal F}\over \partial h }
= {1 \over {\cal A}}  { \partial \Delta {\cal F}_1\over \partial h }
+ {1 \over {\cal A}} { \partial \Delta {\cal F}_2\over \partial h }
+ {1 \over {\cal A}} { \partial \Delta {\cal F}_3\over \partial h },
\en
where ${\cal A}$ is the area of the plane and
\begin{eqnarray}
{1 \over {\cal A}} { \partial \beta \Delta {\cal F}_1\over \partial h } &=&
\int {d^{2}{\bf q} \over (2 \pi)^2} \int_0^h { dz \over 2 \ell_B}\,G_<(z,z,q)
{\partial \Lambda^2\over \partial h } \nonumber \\
{1 \over {\cal A}} { \partial \beta \Delta {\cal F}_2\over \partial h } &=&
\int {d^{2}{\bf q} \over (2\pi)^2} \int_h^{\infty} { dz \over 2 \ell_B}\,G_>(z,z,q)\,
{\partial \over \partial h }\left[ {2 \over ( z- h +\lambda)^2 }\right], \nonumber \\
{1 \over {\cal A}} { \partial \beta \Delta {\cal F}_3\over \partial h } &=&
\left ( \Lambda^2 - { 2 \over \lambda^2 } \right ) \int {d^{2}{\bf q} \over (2\pi)^2}\,
{  G(h,h,q)\over 2 \ell_B} .
\end{eqnarray}
After lengthy algebra, we have\begin{eqnarray}
{1 \over {\cal A}} { \partial \Delta {\cal F}\over \partial h }
= {\Lambda^2 \over h } \left [\,{\cal I}_1(h) - {\cal I}_2(h)\,\right ]
+ { 1 \over \lambda^2 h }{\partial \lambda \over \partial h}
\left [\,{\cal I}_3(h) - {\cal I}_4(h) + 2\,{\cal I}_2(h)\, \right ]
+ { {\cal I}_5(h) \over h^3},
\end{eqnarray}
where the dimensionaless integrals ${\cal I}_i(h)$ are defined as
\begin{eqnarray}
{\cal I}_1(h) &\equiv& h\,\int {d^{2}{\bf q} \over (2\pi)^2} {1 \over 4\alpha} \left \{
{ {\cal M}_2(q) e^{-2 \alpha h}\over 1+ {\cal M}_1(q) {\cal M}_2(q)} -
{ 1- f_+(q) \over \left [1+ f_+(q) \right ]\left [ 1+ {\cal M}_1(q) {\cal M}_2(q) \right ]}\right \},  \\
{\cal I}_2(h) &\equiv&  \int {d^{2}{\bf q} \over (2\pi)^2}
{1 \over 8\alpha^2} {1 - e^{-2 \alpha h}\over 1+ {\cal M}_1(q) {\cal M}_2(q)}
\left [ {\cal M}_2(q)  - { 1- f_+(q) \over 1+ f_+(q)} \right ],  \\
{\cal I}_3(h) &\equiv& h\,\int {d^{2}{\bf q} \over (2\pi)^2} {1 \over 2 \alpha}
\left [ 1- {\alpha  \over q} - {2\,  {\cal M}_1(q) {\cal M}_2(q) \over 1+ {\cal M}_1(q) {\cal M}_2(q) }
\right ],  \\
{\cal I}_4(h) &\equiv& h\, \int {d^{2}{\bf q} \over (2\pi)^2}{ 1 \over 2 q}
{ 1 + \lambda ( \alpha - q) + {\cal M}_2(q) e^{-2 \alpha h} \left [ 1 + \lambda ( \alpha +q) \right ]
\over \left [ 1+ q\lambda + q^2 \lambda^2 + \alpha \lambda ( 1 + q
\lambda )\right ] \left [ 1 + {\cal M}_1(q) {\cal M}_2(q) \right ] } \\
{\cal I}_5(h) &\equiv&  { h^3 \over 2} \int {d^{2}{\bf q} \over (2\pi)^2}
{ ( \alpha - q ) \left [ 1 -  e^{-2 \alpha h}\right ] \over
\left [ 1+ q\lambda + q^2 \lambda^2 + \alpha \lambda ( 1 + q
\lambda )\right ] \left [ 1 + {\cal M}_1(q) {\cal M}_2(q) \right ] }.
\end{eqnarray}
\end{widetext}


\begin{thebibliography}{0}

\bibitem{review1}
A.Yu.Groberg, T.T. Nguyen, and B.I. Shklovskii, Rev. Mod. Phys. {\bf 74}, 329 (2002).

\bibitem{review2}
Yan Levin, Rep. Prog. Phys. {\bf 65}, 1577 (2002).

\bibitem{review3}
A.G. Moreira and Roland R. Netz, in
{\em Electrostatic Effects in Soft Matter and Biophyiscs}, edited
by C. Holm, P. Kekicheff, and R. Podgornik (Kluwer Acad. Pub., Boston, 2001).


\bibitem{contactvalue}
J.N. Israelachvili, {\em Intermolecular and Surface Forces.}
(Academic Press Inc., San Diego, 1992).

\bibitem{fluct1}
B.-Y Ha and A.J. Liu, Phys. Rev. Lett. {\bf 79}, 1289 (1997);
Phys. Rev. Lett. {\bf 81}, 1011 (1998); Phys. Rev. E {\bf 58}, 6281
(1998); Phys. Rev. E {\bf 60}, 803 (1999).

\bibitem{fluct2}
P. Pincus and S.A. Safran, Europhys. Lett. {\bf 42} 103 (1998);
D.B. Lukatsky and S.A. Safran, Phys. Rev. E {\bf 60}, 5848 (1999).

\bibitem{andy}
A.W.C. Lau, P. Pincus, \textit{Phys. Rev. E} {\bfseries 66}, 041501 (2002).

\bibitem{ramin}
R. Golestanian, M. Kardar, and T.B. Liverpool, Phys. Rev. Lett. {\bf 82}, 4456 (1999);
R. Golestanian, T.B. Liverpool, Phys. Rev. E {\bf 66}, 051802 (2002).

\bibitem{schiessel}
H. Schiessel and P. Pincus, Macromolecules {\bf 31}, 7953 (1998).


\bibitem{brushtheory1}
S.J. Miklavic, S. Marcelja, J. Phys. Chem. {\bf 92}, 6718 (1988).

\bibitem{brushtheory2}
S. Misra, S. Varanasi, P.P. Varanasi, Macromolecules, {\bf 22} 5173 (1989).

\bibitem{fylbrush}
P. Pincus, \textit{Macromolecules} {\bfseries 24}, 2912 (1991).

\bibitem{fylpoor}
R.S. Ross, P. Pincus, \textit{Macromolecules} {\bfseries 25}, 2177 (1992);
E.B. Zhulina, T.M. Birshtein, and O.V. Borisov, J. Phys. II {\bf 2}, 63 (1992).

\bibitem{brushtheory3} O.V. Borisov, T.M. Birshtein, and E.B. Zhulina,
J. Phys. II {\bf 1}, 521 (1991); R. Israels, F.A.M. Leermakers, G.J. Fleer,
E.B. Zhulina, Macromolecules {\bf 27}, 3249 (1994); O.V. Borisov, E.B. Zhulina, T.M. Birshtein,
\textit{Macromolecules} {\bfseries 27}, 4795 (1994);

\bibitem{brushtheory3.5}
E.B. Zhulina, O.V. Borisov, \textit{J. Chem. Phys.} {\bfseries 107}, 5952 (1997); E.B. Zhulina,
J. Klein Wolterink, O.V. Borisov, \textit{Macromolecules} {\bfseries 33}, 4945 (2000).


\bibitem{brushtheory4}
J. Wittmer, J.-F. Joanny, Macromolecules {\bf 26}, 2691 (1993).


\bibitem{jlb}
C. Amiel, M. Sikka, J.W. Schneider, Jr., Y.-H. Tsao, M. Tirrell, and  J.W. Mays,
Macromolecules {\bf 28}, 3125 (1995); T.W. Kelley, P.A. Schorr, K.P. Johnson, M. Tirrell,
C.D. Frisbie, Macromolecules {\bf 31}, 4297 (1998)

\bibitem{brush-expt1}
Y. Mir, P. Auvroy, L. Auvray, Phys. Rev. Lett. {\bf 75}, 2863 (1995).

\bibitem{brush-expt2}
P. Guenoun, A. Schlachli, D. Sentenac, J.M. Mays, J.J. Benattar, Phys. Rev. Lett. {\bf 74}, 3628 (1995).

\bibitem{brush-expt3}
H. Ahrens, S. Forster, C.A. Helm, Macromolecules {\bf 30}, 8447 (1997);
Phys. Rev. Lett. {\bf 81} 4172 (1998).

\bibitem{alexdegennes}
S. Alexander, J. Phys. (France) {\bf 38} 983 (1997); P.G. de Gennes, Macromolecules {\bf 13}, 1069 (1980).


\bibitem{tirrell} M.N. Tamashiro, E. Hernandez-Zapata, P.A. Schorr,
M. Balastre, M. Tirrell, P. Pincus, {\textit J. Chem. Phys.} {\bfseries 115} 1960 (2001);
M. Balastre, F. Li, P. Schorr, J. Yang, J. Mays, M. Tirrell, \textit{Macromolecules} {\bfseries 35} 9480 (2002).

\bibitem{Schorr} P.A. Schorr, Ph.D. Thesis, University of Minnesota, 2000.

\bibitem{denis} D. Bendejacq, Ph.D. Thesis, University of Paris, 2002.

\bibitem{seidel}
F.S. Csajka, C. Seidel, \textit{Macromolecules} {\bfseries 4}, 505 (2001);
C. Seidel, Macromolecules {\bf 36}, 2536 (2003).

\bibitem{scalingtheory}
F.S. Csajka, R.R. Netz, C. Seidel, J.-F. Joanny,
\textit{Eur. Phys. J. E} {\bfseries 4}, 505 (2001).

\bibitem{landau}
L.D. Landau and E.M. Lifshitz, {\em Statistical Physics},
(Pergamon, New York, 1980), 3rd ed., rev. and enl. by
E.M. Lifshitz and L.P. Pitaevskii.

\bibitem{condensation}
A.W.C. Lau, D.B. Lukatsky, P.Pincus, and S. Safran, Phys. Rev. E {\bf 5}, 051502 (2002).

\bibitem{netz} R.R. Netz, H. Orland, \textit{Eur. Phys. J. E} {\bfseries 1}, 67 (2000).

\bibitem{deGennes}
de Gennes, P.-G. {\em Scaling Concepts in Polymer Physics}; Cornell University Press: Ithaca, NY, 1979.

\bibitem{milner}
S.T. Milner, T.A. Witten, M.E. Cates, Eurphys. Lett. {\bf 5}, 413 (1988);
Macromolecules {\bf 21}, 610 (1988); S.T. Milner, Science {\bf 251}, 905 (1991).

\bibitem{borisov}
A.M. Skvortsov, I.V. Pavlushkov, A.A. Gorbunov, Y.B. Zhulina,
O.V. Borisov, V.A. Pryamitsyn, Polym. Sci. {\bf 30} 17, 1706 (1988).

\bibitem{notran}
A. Halperin, J. Phys. (France) {\bf 49}, 547 (1988);
Y.B. Zhulina, V.A. Pryamitsyn, O.V. Borisov, Polym. Sci. {\bf 31} 205 (1989);
E.B. Zhulina, O.V. Borisov, V.A. Pryamitsyn, T.M. Birshtein,
\textit{Macromolecules} {\bfseries 24}, 140 (1991); D.R.M. Williams,
J. Phys. II (France) {\bf 3}, 1313 (1993).

\bibitem{arfken}
G. Arfken, {\em Mathematical Methods for Physicists} (Academic Press, San Diego, 1996).

\end{thebibliography}
\end{document}